\documentclass[10pt]{article}
\usepackage{latexsym,amssymb,amsmath}
\usepackage[]{graphicx} 

\def\d{\delta}

\def\e{{\epsilon}}

\newcommand\qed{{\unskip\nobreak\hfil\penalty50\hskip2em\vadjust{}
    \nobreak\hfil$\Box$\parfillskip=0pt\finalhyphendemerits=0\par}}

\newtheorem{theorem}{Theorem}[section]

\newtheorem{definition}{Definition}[section]
\newtheorem{proposition}{Proposition}[section]

\newtheorem{remark}{Remark}[section]

\newcommand{\R}{{\mathbb R}}

\newcommand{\cM}{{\mathcal M}}

\newcommand{\<}{\langle}
\renewcommand{\>}{\rangle}
\renewcommand{\a}{\alpha}
\renewcommand{\b}{\beta}
\renewcommand{\d}{\delta}
\newcommand{\D}{\partial}
\renewcommand{\e}{\varepsilon}

\newcommand{\g}{\gamma}
\newcommand{\G}{\Gamma}

\renewcommand{\l}{\lambda}

\newcommand{\s}{\sigma}

\renewcommand{\t}{\tau}

\newcommand{\z}{\zeta}
\newcommand{\x}{\xi}
\renewcommand{\i}{\infty}
\newcommand{\vp}{\varphi}

\newcommand{\w}{\omega}

\newcommand{\cR}{{\mathcal R}}

\newcommand{\bP}{{\mathbb P}}
\newcommand{\cO}{{\mathcal O}}
\newcommand{\cW}{{\mathcal W}}
\newcommand{\hv}{\hat{v}}
\newcommand{\sech }{\mbox{sech}}
\newcommand{\cJ}{{\mathcal J}}

\title{Solitary waves of the regularized short pulse and Ostrovsky equations}

\author{Nicola Costanzino\thanks{Department of Mathematics, Pennsylvania State University, University Park, PA, 16802. ({\tt costanzi@math.psu.edu}).}
        \and Vahagn Manukian \thanks{Department of Mathematics, North Carolina State University, Raleigh, NC, 27695.
\;\;          Present Address: Department of Mathematics, University of Kansas, Lawrence, KS, 66045 ({\tt
        manukian@math.ku.edu})} 
          \and Christopher K.R.T. Jones \thanks{Department of Mathematics,
University of North Carolina, Chapel Hill, NC, 27599. ({\tt
ckrtj@email.unc.edu})}}

\begin{document}

\maketitle

\begin{abstract}
We derive a model for the propagation of short pulses in nonlinear
media.  The model is a higher order regularization of the short
pulse equation (SPE). The regularization term arises as the next
term in the expansion of the susceptibility in derivation of the
SPE. Without the regularization term there do not exist
traveling pulses in the class of piecewise smooth functions with one discontinuity. However,
when the regularization term is added we show, for a particular
parameter regime, that the equation supports smooth traveling waves which
have structure similar to solitary waves of the modified KdV
equation. The existence of such traveling pulses is proved via the
Fenichel theory for singularly perturbed systems and
a Melnikov type transversality calculation. Corresponding statements
for the Ostrovsky equations are also included.
\end{abstract}

\pagestyle{myheadings} \thispagestyle{plain} \markboth{N.
COSTANZINO, V. MANUKIAN, C.K.R.T. JONES}{SOLITARY WAVES OF THE RSPE
AND OSTROVSKY EQUATIONS}

\section{Introduction}
The propagation of pulses in a dielectric medium is normally modeled
by the nonlinear Schr\"odinger equation (NLSE).  There are two main
assumptions in the derivation of the NLSE.  The first assumption is
that the response of the material to electromagnetic excitation
attains a quasi-steady state, the second assumption being that the
pulse width is large in comparison to the scale of oscillation of
the carrier frequency \cite{NM}.  The experimental generation of
shorter and shorter pulses means that the second assumption in the
derivation of the NLSE may not be satisfied. Indeed it is possible
to generate pulses whose width is just a few cycles of the carrier
frequency \cite{KNNSMSY}. Therefore, new models are needed for short
pulse propagation.  One approach to the propagation of ultra-short
optical pulses is that taken by T. Sch\"afer and C.E. Wayne in
\cite{SW}. There the authors use the full one dimensional Maxwell
equations and formally derive a reduced model by assuming an
experimentally determined optical susceptibility coupled with an
ansatz that captures short pulses. The equation they derive is the
short-pulse equation (SPE)
\begin{equation}
\label{SPE}\partial_{tx} u + \a u + \partial_x^2 u^3 = 0
\end{equation} where $u : (t,x) \in \R_+ \times \R \rightarrow \R$ is the real component of the electric field in the direction transverse to the
direction of
propagation and $\a \in \R$ is a constant. It was shown there that
\eqref{SPE} does not support traveling waves that are $C^2$ smooth.
They then propose that this might reflect the Maxwell equations not possessing traveling wave
solutions in the short-pulse regime or that equation \eqref{SPE} is an accurate model only for
short times, and the breakdown of solutions occurs after the time of
validity of \eqref{SPE} as an approximation to the Maxwell
equations. We consider here another approach, namely that such traveling pulses exist if higher order terms in the approximation to the
susceptibility are accounted for.

First, we strengthen the non-existence result of \cite{SW} and show there are not even piecewise smooth traveling pulses to \eqref{SPE} with a single peak.
The possibility of piecewise smooth traveling waves is motivated by
the development of discontinuities for the Cauchy problem for
\eqref{SPE}. Typically, one cannot construct global smooth solutions
to hyperbolic equations such as \eqref{SPE} since discontinuities in
$\partial_t u$ and $\partial_x u$ tend to develop in finite time.
Since traveling waves are globally-in-time defined objects, the
development of these discontinuities is a definite obstruction to
the existence of smooth traveling waves. Therefore it is natural to
look for traveling waves whose profiles $u(z)$ are smooth on either
side of discontinuity occurring at some point $z=z_\ast$, and at
$z_\ast$ an algebraic condition is satisfied so as to ensure the
piecewise smooth function is a distributional solution to
\eqref{SPE}.  In section \ref{sect:genprop_RSPE} we show that
the SPE does not support traveling pulses in the space of piecewise
smooth functions so that scenario (iii) above is also not possible.
 As explained in Section \ref{sect:genprop_RSPE}, the main
obstruction to the existence of traveling waves is the discontinuity
in the right-hand side of the profile equations
\eqref{SPE-profile-equations}. Indeed, the discontinuity restricts
homoclinic solutions to lie in the set $-\sqrt{c/3} < u < +
\sqrt{c/3}$, whereas the jump condition requires $c= u_\ast^2$ where
$c$ is the speed of the traveling wave and $u_\ast$ is the absolute
value of $u$ immediately to the right (or equivalently left) of the
discontinuity. Clearly these two conditions are not compatible. We
resolve this issue by proposing a regularization mechanism which has
the effect of removing the discontinuity in the profile equations as
well as the need for a jump condition. The nonexistence of traveling
waves (whether smooth or with discontinuities) leads us to
reinvestigate the derivation of the SPE.  By approximating the
linear response function $\chi^{(1)}$ of the dielectric media to
higher order we arrive at the regularized short pulse equation
(RPSE)
\begin{equation}
\label{RSPE}
\partial_x
\partial_t u + \a u +
\partial_x^2 u^3 + \b \partial_x^4 u = 0 \, .
\end{equation} which is the SPE \eqref{SPE} with a higher order dispersive term
of strength $\beta$.

We remark here that the RSPE is similar to the Ostrovsky equation
\begin{equation}
\label{ostrovsky}
\partial_x \left( \partial_t u + \partial_x u^2 + \b \partial_x^3 u
\right) + \a u =0
\end{equation} which was derived by L.A. Ostrovsky \cite{O}
as a model for internal solitary waves in the ocean with rotation
effects of strength $\a<0$. The main differences being the cubic
nonlinear term and both the signs and size of the physical
parameters.\footnote {We note that the regularized short pulse
equation was proposed earlier in \cite{NSC} in the context of plasma
physics. In that context the equation make sense for any combination
of parameters signs.} If $\beta$ is small \eqref{RSPE} is a
singularly perturbed equation which induces a fast-slow structure in
the equations that can be exploited by relating certain rescaled
versions of the equation to other well known equations. Indeed, by
setting $\t = t/\sqrt{|\b|}$ and $\xi = x/\sqrt{|\b|}$, one may
rewrite \eqref{RSPE} as
\begin{equation}
\label{rescaled_RSPE}
\partial_\x \partial_\t u  +  \a |\b| u + \partial_\x^2 u^3 +{\rm sgn}(\b) \partial_\x^4 u =
0 \, .
\end{equation} Setting $\b = 0$ in the RSPE with the usual scaling \eqref{RSPE} yields the
SPE while setting $\b = 0$ in the fast scaling \eqref{rescaled_RSPE}
yields the modified Korteweg deVries equation (mKdV)
\begin{equation}
\label{mKdV}
\partial_t u + \partial_x u^3 + {\rm sgn}(\b) \partial_x^3 u = 0 \, .
\end{equation}  Hence in the slow scaling the RSPE is a small perturbation of the SPE while in the fast scaling the RSPE is a small perturbation of
the mKdV equation.  Given this, we expect that the traveling waves $u(x- c t)$ of the
RSPE, if they indeed do exist, are close to solutions of the SPE for
$|x- c t| \sim {\mathcal O}(1/\sqrt{|\b|}) $, and close to traveling
waves of mKdV for $|x - c t| \sim {\mathcal O}(\sqrt{|\b|})$ with a
transition region for $|x-c t| \sim {\mathcal O}(\sqrt{|\b|})$. This
scenario is proven via geometric singular perturbation theory in
section \ref{sect:existence_RSPE}. While in the derivation of the
RSPE $\beta$ need not be small, one can nevertheless introduce a
small parameter $\e$ \eqref{epsilon} whereby under a suitable
rescaling the equation is singularly perturbed in $\e$.

\section{Derivation of the Regularized Short Pulse Equation}
\label{sect:derivation} Consider the  Maxwell equations in three
space dimensions,
\begin{align}
\label{maxwell}
\begin{split}
\partial_t{\bf B} = -\nabla\times {\bf E}, \qquad \nabla\cdot {\bf D} =
\rho  \\
\partial_t {\bf D} = \nabla\times {\bf H}
- {\bf j}, \qquad \nabla \cdot {\bf B} = 0
\end{split}
\end{align}
where ${\bf E}, {\bf H} : (x,t) \in \R^3 \times \R_+ \mapsto \R^3$
are the electric and magnetic fields, ${\bf D}, {\bf B} : (x,t) \in
\R^3 \times \R_+ \mapsto \R^3$ are the electric and magnetic flux
densities, ${\bf \rho} \in \R$  is the electric charge and ${\bf j}
: (x,t) \in \R^3 \times \R_+ \mapsto \R^3$ is the current density.
To derive \eqref{RSPE}, we make several assumptions about the
physical setup. The first is that the medium is a dielectric. This
implies that the there are no free charges or currents.  The second
is that the dielectric medium is cubic and independent of space.
These assumption are written as
\newline

\noindent (H1) $\hspace{0.5cm} \rho \equiv 0 \;\;\; \mbox{and} \;\;\; \bf{j}= 0$ \\

\noindent (H2) $\hspace{0.5cm}   {\bf B} = \mu_0 {\bf H} \;\;\; \mbox{and} \;\;\; {\bf D} = \e_0 {\bf E} + \e_0{\bf P}(x,y,z,{\bf E})$\\

\noindent (H3) $\hspace{0.5cm} {\bf P}(x,y,z,{\bf E}) = \chi^{(1)} \ast {\bf E} + \chi^{(3)} \ast |{\bf E}|^2 {\bf E}$\\

\noindent where $\mu_0, \e_0$ are free space constants and $\ast$
denotes convolution in time t. Hence, Maxwell equations become

\begin{align}
\label{maxwell_2}
\begin{split}
(a)& \hspace{1.0cm} \e_0 \partial_t {\bf E}  = \nabla \times {\bf H} - \e_0 \partial_t  \left\{ \chi^{(1)} \ast {\bf E} +   \chi^{(3)} \ast |{\bf
E}|^2 {\bf E} \right\} \\
(b)& \hspace{1.0cm}\mu_0 \partial_t {\bf H} = - \nabla \times {\bf E}
\end{split}
\end{align}  We turn \eqref{maxwell_2} into a wave equation for ${\bf E}$, by
combining the time derivative of \eqref{maxwell_2}(a), with
\eqref{maxwell_2}(b). Recalling the identity $\nabla \times \nabla
\times {\bf E} = \nabla (\nabla \cdot {\bf E}) - \nabla^2 {\bf E}$,
we get
\begin{equation}
\label{fiber} \nabla^2 {\bf E}  - \nabla (\nabla \cdot {\bf E})=
\mu_0 \e_0 \partial_t^2 \left( {\bf E} + \chi^{(1)} \ast {\bf E} +
\chi^{(3)} \ast |{\bf E}|^2 {\bf E} \right).
\end{equation} We look for solutions of the form
\begin{equation}
\label{ansatz} {\bf E}(x,y,z,t) = v(x,t){\bf E}^{\perp}(y,z,t)
\end{equation}  where $v : \R \times \R_+ \rightarrow \R$ and ${\bf E}^\perp : \R^2 \times \R_+ \rightarrow \R^3$ is given by
\[
{\bf E}^\perp = (0,E_2,E_3)
\]  Substituting \eqref{ansatz} into \eqref{fiber} yields
\begin{align}
\begin{split}
{\bf E}^{\perp}\left( \partial_x^2 v - \mu_0 \e_0 \partial_t^2 v \right) - \mu_0 \e_0 \left( 2 \partial_t v \partial_t {\bf E}^\perp + v \partial_t^2
{\bf E}^\perp    \right) + v \nabla^2 {\bf E}^\perp - \nabla \left( v \nabla \cdot {\bf E}^\perp \right) \\
= \mu_0 \e_0 \partial_t^2 \left( \chi^{(1)} \ast (v {\bf E}^{\perp})
+ \chi^{(3)} \ast (v^3 |{\bf E}^{\perp}|^2{\bf E}^{\perp}) \right).
\end{split}
\end{align} Since we are interested in the evolution of ${\bf E}$ along $x$, we
can start by looking for solutions for which ${\bf
E}^{\perp}(y,z,t)$ is constant vector. In this case, by simply
redefining the $\chi^{(j)}$ by constants we find that the equation
for the transverse evolution of the electromagnetic field satisfies
the scalar one dimensional equation
\begin{equation}
\partial_x^2 v  - \mu_0 \e_0 \partial_t^2 v = \mu_0 \e_0 \partial_t^2 \left( \chi^{(1)} \ast v + \chi^{(3)} \ast v^3
\right) \,.
\end{equation} Here we view $x$ as the evolution variable, so to
put this in a form in which $t$ is the evolution variable we make
the change of coordinates $(x,t) \mapsto (t,x)$ and set $(\mu_0
\e_0)^{-1} =: c_0^2$ to get
\begin{equation}
\label{nonlinear_maxwell} \D_t^2 v - \frac{1}{c_0^2}\D_x^2 v =
\frac{1}{c_0^2}
\partial_x^2 \left(\chi^{(1)} \ast v + \chi^{(3)} \ast v^3 \right)
\end{equation} where now $\ast$ denotes convolution with respect to $x$.  It is
convenient to consider the Fourier transform of
\eqref{nonlinear_maxwell} which we write as
\begin{equation}\label{FT_nonlinear_maxwell}
\D_t^2 \hv + \frac{\xi^2}{c_0^2} \hv := -\frac{\xi^2}{c_0^2}
\left(\hat{\chi}^{(1)}\hv + \hat{\chi}^{(3)}\widehat{v^3} \right)
\,.
\end{equation} \\

\begin{remark}Since we are interested primarily in modeling the effect of the $\chi^{(1)}$ term, for simplicity we will assume that the $\chi^{(3)}$
response is
instantaneous and modeled by the Dirac measure.  Thus in
\eqref{FT_nonlinear_maxwell} and what follows we set
$\hat{\chi}^{(3)} = 1$. We investigate the effect of different
$\chi^{(3)}$ response functions in future work.
\end{remark}

\subsection{Approximating the $\chi^{(1)}$ Term}  In dielectric
materials, the $\chi^{(1)}$ term is modeled in Fourier space as
\cite{B,NM}
\begin{equation}
\label{chi_1} \hat{\chi}^{(1)}(\xi) = \sum_{j=1}^n a_j \left(
\frac{2\xi_j}{\xi_j^2 - \xi^2 + \d_j^2 + 2i\d_j \xi} \right).
\end{equation}
Typically for silica fibers, there are three resonances of
importance which occur at practical wavelengths, namely at
wavelengths of $\l = 0.068 \mu m, \l = 0.116 \mu m$ and $\l = 9.896
\mu m$. If one restricts attention to wavelengths between 0.25 and
3.5 $\mu m$, approximate values for the various constants in
\eqref{chi_1} can be obtained by fitting experimental data for light
propagation in silica to obtain \cite{M}

\begin{equation}\label{real_approx}
\hat{\chi}^{(1)}_{\rm silica}(\l) = \frac{0.696 \l^2}{\l^2 -
(0.0684)^2} + \frac{0.4079 \l^2}{\l^2 - (0.116)^2} + \frac{0.8974
\l^2}{\l^2 - (9.896)^2}.
\end{equation} where $\l$ is the wavelength
\begin{equation}\label{lambda}
\l = \frac{2\pi}{\xi}.
\end{equation}

The idea in \cite{SW}, developed further in this section, is that
\eqref{real_approx} can be expanded in powers of $\lambda$, and that
the expansion is in fact a good approximation over a particular
range of wavelengths. Indeed, for wavelengths between $1.6 \mu m$
and $3.0 \mu m$ one can approximate \eqref{real_approx} by
\begin{equation}
\label{chi_1_approx} \hat{\chi}_{\rm
silica}^{(1)\;\tilde{\a},\tilde{\mu} }(\l) = \tilde \mu - \tilde \a
\left( \frac{\l}{2\pi} \right)^2
\end{equation} for appropriate values of $\tilde \a$ and $\tilde \mu$ (see \cite{SW,CJSW} for a discussion on the validity of this
approximation). Here we write an expansion of $\hat{\chi}^{(1)}$
beyond that of \cite{SW} by adding an additional term,
\begin{align}
\label{our_approx}
\begin{split}
 \hat{\chi}_{\rm
silica}^{(1)\;\tilde{\a},\tilde{\mu}, \tilde{\nu}}(\l) &
=\hat{\chi}_{\rm silica}^{(1)\;\tilde{\a},\tilde{\mu} }(\l) - \tilde
\nu \left( \frac{2\pi}{\l} \right)^2 \\
& = \tilde \mu - \tilde{\a}\left(\frac{\l}{2\pi} \right)^2 - \tilde
\nu \left( \frac{2\pi}{\l} \right)^2 \, .
\end{split}
\end{align}  For any fixed range of wavelengths, \eqref{our_approx} leads to a
better approximation to \eqref{real_approx} since it contains three
free parameters $\tilde{\a}, \tilde{\mu}, \tilde{\nu}$ instead of
just two as in \eqref{chi_1_approx}.
Note also that since $\lambda = 2\pi/ \xi$, the
additional $\tilde{\nu}$ term corresponds to a higher order
derivative in physical space. 
Finally, due to the convolution term in the nonlinear
Maxwell equation (and hence the product of $\hat{\chi}^{(1)}$ and
$\hat{u}$ in \eqref{FT_nonlinear_maxwell}), an implicit assumption
in the derivation is that $\hat u$ is small outside the range for
which $\chi^{(1)}$ is well approximated.

Recalling \eqref{lambda} and plugging the expansion for $
\hat{\chi}_{\rm silica}^{(1)\;\tilde{\a},\tilde{\mu},
\tilde{\nu}}(\xi)$ \eqref{our_approx} into
\eqref{FT_nonlinear_maxwell} and transforming back to physical
variables leads to
\begin{align}
\label{quasi}
\partial_t^2 v - \g^2 \partial_x^2 v = \a v +
\partial_x^2 v^3 + \nu \partial_x^4 v
\end{align} where
\begin{align}\label{physical-parameters}
\a = \tilde \a /c_0^2, \;\; \mu = \tilde \mu /c_0^2, \;\;  \nu =
\tilde \nu/c_0^2, \;\; \g^2 = \frac{1+ \tilde \mu}{c_0^2} \, .
\end{align}

\subsection{Parameter Values}\label{sect:par_vals}
Up to this point we have made no mention of the signs
or actual values of the parameters in \eqref{our_approx}. In this section we compute the signs and actual values of the
parameters $\tilde{\alpha}, \tilde{\mu}$, and $\tilde{\nu}$ which
depend on the wavelength regime over which the $\hat{\chi}^{(1)}$
response function \eqref{real_approx} is approximated. We follow the work of Sch\" afer and Wayne \cite{SW}, in making a least-squares fit to \eqref{real_approx} over the
range $1.6 \mu m \leq \l \leq 3.0 \mu m$, see Fig 2.1. The sign of the parameter $\tilde{\nu}$ in \eqref{our_approx} leads to a parameter regime for which our results hold. This approximation does not capture the behavior of the response function over a larger region, in particular one that includes its upwardly convex part when $\lambda < 1$. A least-squares fit over the wavelength range $0.25 \mu m \leq \lambda \leq 3.5 \mu m$ gives a value of $\tilde{\nu}$ with the opposite sign. Our analysis does not cover this case, but we should point out that, although the approximation is better over the larger frequency range, it is not as good as the one in \cite{SW}, which we use here, near the carrier frequency of interest.

\begin{figure}[!h] \label{fig_tobias}
\begin{center}
\includegraphics[angle=0, width=1.00\textwidth]{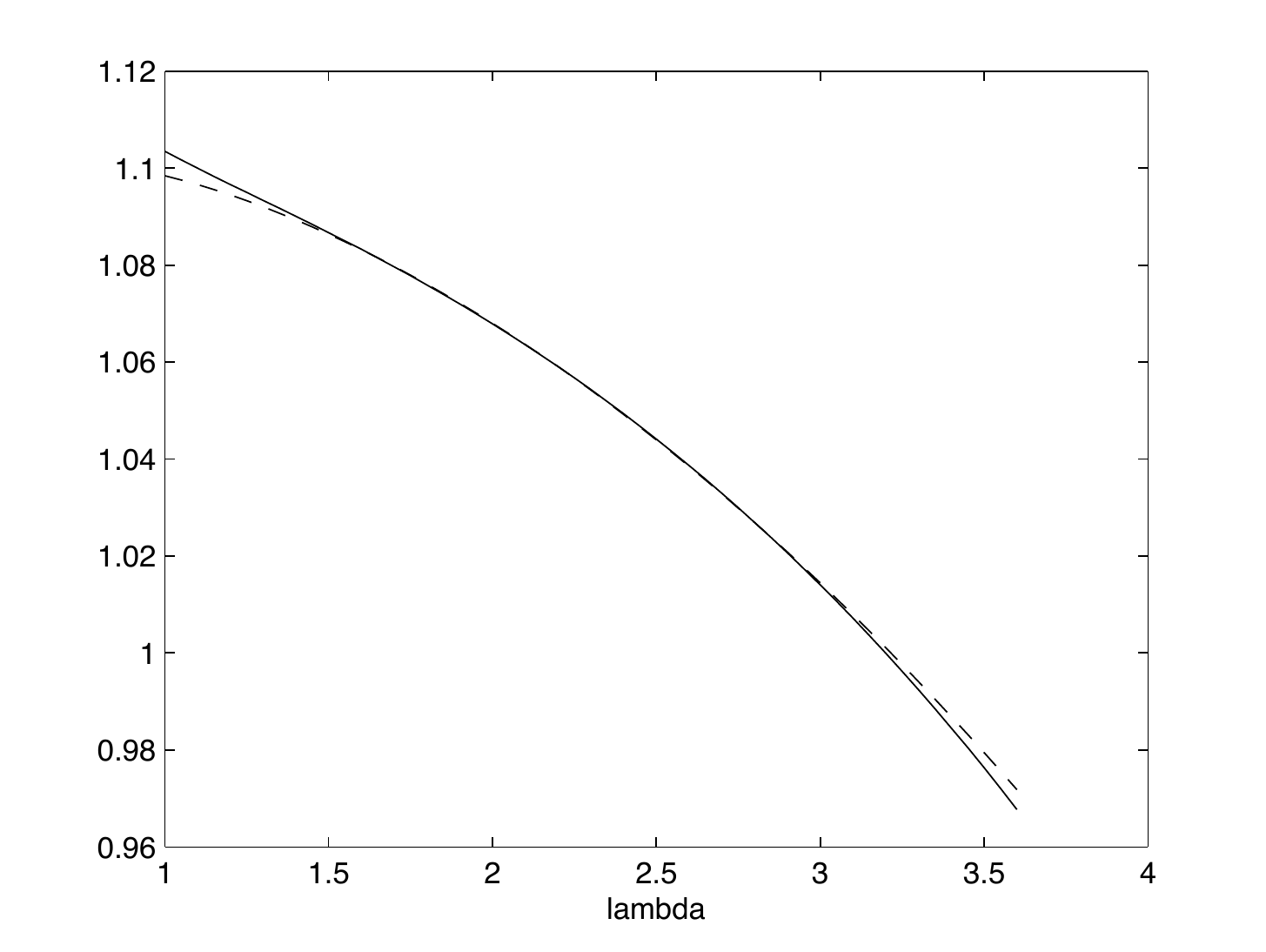}
\caption{Approximation of the response function for silica using
\eqref{our_approx} with $\tilde{\alpha} = 4.287786709 \times 10^{-1}, \;\; \tilde{\mu} =
1.112628599, \;\; \tilde{\nu} = 1.192774270 \times 10^{-4}$.  The dashed line is \eqref{our_approx},
the solid line is \eqref{real_approx}.}
\end{center}
\end{figure}

\subsubsection{Approximation over the regime $1.6 \mu m \leq \lambda \leq 3.0 \mu
m$} \label{sect:small_regime} Here we compute the parameters
$\tilde{\a},\tilde{\mu}, \tilde{\nu}$ for the ansatz
\eqref{our_approx} by curve fitting over the exact same wavelength
range as in the original paper \cite{SW}. The idea is that the extra
free parameter $ \tilde{\nu}$ leads to a better fit, and in addition
provides a non-zero higher order regularization term. In this case
the parameters obtained by a least-squares fit are

\begin{align}
\tilde{\alpha} = 4.287786709 \times 10^{-1}, \;\; \tilde{\mu} =
1.112628599, \;\; \tilde{\nu} = 1.192774270 \times 10^{-4} \; .
\end{align}

In light of \eqref{physical-parameters} this leads to an RSPE
equation \eqref{RSPE} with $\alpha>0, \beta>0$.

\subsection{The Short Pulse Ansatz} To get \eqref{SPE} and
\eqref{RSPE} from \eqref{quasi} we introduce the short pulse
expansion
\begin{equation}
\label{short-pulse-ansatz}
 v(x,t) = \d v_0(x^\d,t^\d) + \d^2 v_1(x^\d,t^\d) +
 \cdot\cdot \cdot
\end{equation} where
\begin{equation}
x^\d = \frac{x-\g t}{\d} \hspace{1.0cm} t^\d = \frac {\d}{2}t.
\end{equation}  Let
\begin{align}
\nu = \beta \delta^4, \;\;\; \d \ll \b
\end{align} then substitution of \eqref{short-pulse-ansatz} into \eqref{quasi} yields
\begin{equation}\label{der4}
\D_{tx}v_0 + \a v_0 + \D_{x}^2 v_0^3 + \beta \D_{x}^4 v_0 =
\mathfrak{o}(\d)
\end{equation} where we have dropped the $\d$ superscripts on $x,t$.  Then to solve \eqref{der4} to order $\d$ we need to solve
\begin{equation}
\D_{tx} v_0  + \a v_0 + \D_{x}^2 v_0^3 + \b \D_{x}^4 v_0 = 0 \, .
\end{equation}

\section{General Properties of the RPSE}
\label{sect:genprop_RSPE}

\subsection{Conserved Quantities and Hamiltonian Formulation}
 While both the SPE the mKdV equation are
integrable \cite{AC,SS} so associated with them are an infinite
number of conserved quantities, the arguments in \cite{GGS,CIL} show
that the $\a$ term in the RSPE destroys the integrable structure.
Nevertheless there are conserved quantities associated with
\eqref{RSPE}. These are the zero-mass law
\begin{equation}
\label{conserved-1} m(u) := \int_\R u(x,t) \, dx = 0,
\end{equation} the momentum
\begin{equation}
\label{conserved-2} M(u) := \int_\R u^2 \, dx
\end{equation} and the energy
\begin{equation}
H(u) := \frac{1}{2}\int_\R  \a |\partial_x^{-1} u|^2 - \frac{1}{2}
|u|^4 + \b |\partial_x u|^2 \, dx = const.
\end{equation}  With this definition for $H$, one can regard $H$ as the Hamiltonian for the RSPE since one can write it as the gradient flow
\begin{equation}
\partial_t u = \cJ \nabla H(u)
\end{equation} where $\cJ$ is the skew symmetric operator
\begin{equation}
\cJ = -\partial_x \, .
\end{equation}

\subsection{Dispersive Regimes}
Consider the linear RSPE equation obtained by omitting the nonlinear
term,
\begin{equation}\label{linear_RSPE}
\D_{tx}u + \a u + \b \D_x^4 u = 0 \,.
\end{equation} Plane waves of the form $u = Ae^{i(\x x - \w t)}$ are
solution to \eqref{linear_RSPE} provided that the pair $(\x,\w)$
satisfy the dispersion relation
\begin{align}\label{dispersion-relation}
c(\xi) = \frac{\a}{\x^2} + \b \xi^2
\end{align} where $c(\xi) = \w / \x$ is the phase speed. The dispersion relation \eqref{dispersion-relation} shows us there
are two different dispersion mechanisms. For the sake of
nomenclature, we call the $\a/\x^2$ term {\em lower-order
dispersion} and the $\b \xi^2$ term {\em higher-order dispersion}.
There are essentially four dispersive regimes one can investigate.
When both lower-order and higher-order dispersion mechanisms are
absent ($\a=\b=0$) the RSPE reduces to a Burgers equation with
nonconvex flux which is known to develop shocks (c.f. \cite{Daf}).
When only the lower-order dispersion mechanism is absent, the RSPE
reduces to the modified KdV equation which has solitary wave
solutions (corresponding to homoclinic orbits)
\begin{align}
q^0 = \sqrt{2}\sech\left(\sqrt{\frac{c}{|\b|}}z\right),
\hspace{0.5cm} z = x-ct, \hspace{0.5cm} c,\b > 0
\end{align} and traveling front solutions (corresponding to heteroclinic orbits)
\begin{align}
q^\pm = \pm \sqrt{c} \tanh \left(  \sqrt{\frac{c}{2|\b|}} z \right),
\hspace{0.5cm} z = x-ct, \hspace{0.5cm} c > 0, \b < 0
\end{align} When only the higher order dispersion mechanism is absent, the RSPE
equation reduces to the SPE equation.  Finally, we investigate the
regime where {\em both} dispersion mechanisms play a role, which is
the RSPE equation.
\section{Solitary Waves of the SPE and RSPE}
\subsection{Nonexistence of Traveling Waves for the SPE}\label{sect:nonexistence_SPE}
Here we prove that there do not exist
piecewise smooth traveling pulse solutions to the SPE. By ``pulse'' here we mean that there is only one
point of discontinuity. Rewrite
\eqref{SPE} as the system
\begin{align}
\label{SPE-system}
\begin{split}
& (a) \hspace{1.0cm} \partial_x q = u\\
& (b) \hspace{1.0cm} \partial_t u + \a q + \partial_x u^3 = 0
\end{split}
\end{align} which can be thought of as an integrated version of
\eqref{SPE} in the sense that one can obtain \eqref{SPE} by taking
the spatial derivative of \eqref{SPE-system}(b).  In this sense the
quantity $q$ would represent the {\em electric potential} of $u$.
Written this way the hyperbolic nature of the SPE is apparent in
that one can view it as a correction to Burgers equation with
nonconvex flux.  Given this, we do not expect the existence of
smooth traveling waves for \eqref{SPE} since in general the initial
value problem for equations of this type tend to form
discontinuities in the derivatives $\partial_t u$ and $\partial_x
u$, which means we must look for distributional solutions.  This has
already been proven in \cite{SW}. Therefore, we look for traveling
waves with profiles $(q,u)$ that are smooth on either size of a
point of discontinuity $z_\ast$ which are distributional solutions
to the original PDE \eqref{SPE}. The requirement that the traveling
wave be a distributional solution yields an algebraic condition that
must be satisfied at the discontinuity, which are the usual
Rankine-Hugoniot jump conditions. Traveling waves with these types
of discontinuities are phenomena in both hyperbolic evolution
equations (c.f. \cite{CF,MN}) and parabolic equations (c.f.
\cite{BFRW,Fi,MN}).

It is not hard to see that piecewise smooth functions $(q,u)$ with discontinuities across a
smooth curve ${\mathcal S}:= \{(t,x) : x = \sigma(t)\}$ with unit
normal ${\bf \nu} \equiv (\nu^1, \nu^2) = (1+
\dot{\sigma}^2)^{-1/2}(1,-\dot{\sigma})$ are weak solutions of
\eqref{SPE-system} if and only if \eqref{SPE-system} is satisfied in
the classical sense on either side of ${\mathcal S}$ and the traces
of $(q,u)$ satisfy the jump condition

\begin{align}
\label{RH}
\begin{split}
&(a) \hspace{1.0cm}  [[q]] = 0 \\
&(b) \hspace{1.0cm}  \dot \s [[u]] = [[u^3]]
\end{split}
\end{align} where $[[ \; \cdot\;  ]]$ denotes the jump across the curve ${\mathcal S}$.

Note that \eqref{RH} implies $q$ is continuous across ${\mathcal S}$.
In the case of symmetric Jumps, we can make an explicit calculation.
Let $u_\ast \neq 0$ and suppose $u({\mathcal S}+0) = u_\ast = -
u({\mathcal S}-0)$. Then
\begin{equation}
\dot \sigma = u_\ast^2 > 0.
\end{equation}

Suppose the pair $(q,u)(z)$ is smooth
on either size of $z=0$ with nonequal left- and right-hand
limits at $z=0$.  Then $(q,u)(z)$ is a distributional solution to \eqref{SPE-system} if and only if the following hold \\

\noindent {\bf 1.}  $(q,u)$ satisfy the profile equations

\begin{align}\label{SPE-profile-equations}
\begin{split}
& \dot q = u \\
& \dot u = \frac{\a q}{c - 3u^{2}}
\end{split}
\end{align} pointwise in
$\{ z<0 \}$ and $\{ z>0 \}$ \\

\noindent {\bf 2.} The Rankine-Hugoniot conditions
\begin{equation}
\label{SPE-transmision-equations}
\begin{array}{rcl}
 [[q]] &=& 0 \\
 c[[u]] &=& [[u^3]]
\end{array}
\end{equation} hold on $z=0$.

Now write the profile equations \eqref{SPE-profile-equations} and
transmission condition \eqref{SPE-transmision-equations} as
\begin{equation}
\label{SPE-profile-system}
\begin{array}{rll}
\dot w &= F(w) \hspace{1.0cm} \pm z > 0\\
 c \left[ \left[ w \right] \right] & = \left[ \left[ G(w) \right] \right] \hspace{1.0cm} z = 0
 \end{array}
\end{equation} where
\begin{equation}
w := \left( \begin{array}{c} q \\ u \end{array} \right)
\hspace{1.0cm} F(w) := \left( \begin{array}{c} u \\
\frac{\a q}{c - 3u^{2}}
\end{array} \right)  \hspace{1.0cm} G(w) := \left( \begin{array}{c} 0 \\ u^3 \end{array} \right)
\end{equation}  The origin $0\in\R^2$ is the unique fixed point of \eqref{SPE-profile-system}.
The linearization of $F$ about the origin yields the linear system
\begin{equation}
\dot w = A w
\end{equation} with
\begin{equation}
A = \left( \begin{array}{cc}
0 & 1  \\
\a/c & 0
\end{array} \right) \, .
\end{equation} so that it is a hyperbolic fixed point if
\begin{align}\label{I:existence:a/c>0}
\a/c > 0.
\end{align}

 Since $0 \in \R^2$ is a hyperbolic fixed point of
\eqref{SPE-profile-system}, there exist stable
and unstable manifolds $\cW^s(0)$, $\cW^u(0)$ which locally can be
described as graphs over the stable and unstable eigenspaces $E^s$,
$E^u$ of the linearized system.  These manifolds can be characterized the Hamiltonian structure of the profile
equations. Define $H(q,u)$ by
\begin{equation}
\label{potential} H (q,u) = \a q^2 - c u^2 + \frac{3}{2} u^{4}
\end{equation} Let
\begin{align}
\begin{split}
& (a) \hspace{1.0cm} \Sigma_- = \{(q,u) \in \R^2 :|u| < \sqrt{c /3}, q u \leq 0\}\\
& (b) \hspace{1.0cm} \Sigma_+ =\{(q,u) \in \R^2 : |u| < \sqrt{c /
3}, qu \geq 0\}
\end{split}
\end{align} and define the sets $H_-, H_+$ by
\begin{align}
\label{H-+}
\begin{split}
& (a) \hspace{1.0cm} H_- := \{ H(q,u) = 0 \} \cap \Sigma_- \\
& (b) \hspace{1.0cm} H_+ := \{ H(q,u) = 0 \} \cap \Sigma_+ \\
\end{split}
\end{align}
It is not hard to see that $H_- = {\mathcal W}^s(0)$ and $H_+ = {\mathcal W}^u(0)$.

 Next suppose that
$\overline w(z)$ is a piecewise smooth traveling wave.  Then
$\overline w(z)$ is defined through the stable and unstable
manifolds \eqref{H-+} by
\begin{equation}
\label{peakon} \overline w(z) =  w_-(-z) \chi_{\{z < 0\}} + w_+(z)
\chi_{\{z
> 0\}}
\end{equation} where $w_+$ and $w_-$ are solutions to the initial value problem \eqref{SPE-profile-equations} with initial data
\begin{equation}
w_+^0 = (q_\ast, u_\ast) \in {\mathcal W}^s(0) \hspace{2.0cm} w_-^0=
(q_\ast, -u_\ast) \in {\mathcal W}^s(0)
\end{equation}
and $\chi$ is the characteristic function.  We will show that
$\overline w(z)$ defined by \eqref{peakon} cannot simultaneously
satisfy the jump conditions and at the same time lie within the
stable and unstable manifolds for every $z \in \R$. By construction,
$\overline w(z)$ is smooth away from $z=0$ and satisfies
\eqref{SPE-profile-system} on $\R \slash \{ 0 \}$, thus $\overline
w(z)$ satisfies the profile equation \eqref{SPE-profile-system}
pointwise on $\R \slash \{ 0\}$. Notice that by the Rankine-Hugoniot
jump conditions and the fact that $(q_\ast,u_\ast)$ must lie on
${\mathcal W}^s(0)$, $\overline w$ has symmetric jumps, therefore the speed $c$ of the traveling
wave $\overline w(z)$ is
\begin{equation}
\label{c}
 c = u_\ast^2 >0
\end{equation}  However, for $u_\ast$ to lie in ${\mathcal W}^s(0)$ we
need
\begin{equation}
\frac{c}{3} > u_\ast^{2}
\end{equation} which contradicts
\eqref{c}.\\

\subsection{Existence of Solitary Waves of the RSPE and Ostrovsky Equation}
\label{sect:existence_RSPE} Loosely speaking, the mechanism that
leads to the nonexistence of traveling waves for the SPE is the
discontinuous right-hand side of the profile equations.  As shown in
Section \ref{sect:nonexistence_SPE}, the main obstruction to the
existence of piecewise smooth traveling waves to the SPE is that in
order for the the jump condition to hold, the value of $u$ at the
jump must be larger than the allowable range $|u| < \sqrt{c/3}$
dictated by the singularity in the right-hand side of the profile
equations. If one could eliminate this discontinuity, then it might
be possible to obtain traveling waves.  In order to eliminate the
discontinuous right hand side of the profile equations, we
introduced a small higher order term into the equation.  This higher
order term arose in expanding the susceptibility further so as to
eliminate the zero wavelength singularity inherent in the derivation
of the SPE.  Here we study the regularized short pulse equation
\eqref{RSPE} and show that higher order dispersion is needed to
obtain smooth pulses. For the sake of generality we will prove the
theorem for general pure power nonlinearities,
\begin{align} \label{g-RSPE}
\partial_{tx} u + \a u + \partial_x^2 u^{k} + \b \partial_x^4 u =
0.
\end{align} which includes both the Ostrovsky equation (k=2) and RSPE
(k=3).\\

We now state the main theorem of this section, the proof of which
will occupy the next several subsections.\\

\begin{theorem}[Existence of Traveling Waves for the RSPE] \label{I:theorem:existence_RSPE_waves}
Consider the RSPE with general pure power nonlinearity
\eqref{g-RSPE} and let
\begin{equation}\label{epsilon}
\e = \frac{\sqrt{|\a| |\b|}}{|c|} \,.
\end{equation}

\noindent {\bf 1.} Suppose $k$ is an {\em even} integer and $\e$ is
given by \eqref{epsilon}. Then for $\e
>0$ small enough there exists a traveling wave solution
\begin{equation}
u=u(z), \;\;\; z= x-ct, \;\;\; u(z) \rightarrow 0 \; \mbox{as} \; z
\rightarrow \pm \i
\end{equation} provided that
\begin{equation}
{\rm sgn}(c) = {\rm sgn}(\a) = {\rm sgn}(\b) .
\end{equation}

\noindent {\bf 2.} Suppose $k$ is an {\em odd} integer and $\e$ is
given by \eqref{epsilon}.  Then for $\e>0$ small enough there exists
a traveling wave solution
\begin{equation}
u=u(z), \;\;\; z= x-ct, \;\;\; u(z) \rightarrow 0 \; \mbox{as} \; z
\rightarrow \pm \i
\end{equation} provided that
\begin{equation}
{\rm sgn}(c) = {\rm sgn}(\a) ={\rm sgn}(\b) = +1 .
\end{equation}
\end{theorem}

\begin{remark} Note that when $k$ is an {\em odd} integer, as in the RPSE, we are only able to cover the physically relevant case $\beta>0$ (see
Section \ref{sect:small_regime}). The case $\b < 0$ is
also physically relevant for RSPE (see Section
\ref{sect:par_vals}, $0.25 \mu m \leq \lambda \leq 3.5 \mu m$) but it not considered herein.
\end{remark} 

\begin{remark}  The small parameter $\e$ is a ratio of three parameters and can be thought of as being
made small by fixing two of the parameters and varying one of them.
Hence by this rescaling our analysis of traveling waves of the RSPE
covers the three cases (i) $\a$ small, (ii) $\b$ small, and (iii)
$c$ large.
\end{remark} 

\begin{remark} Similar results can be found in \cite{LL,LV}.  There the
authors use variational arguments to prove the existence of a ground
state traveling wave to the Ostrovsky equation for small speeds $c$.
The GSPT approach used here forms the basis of work in progress \cite{CJMS} on
the construction of multi-pulse and periodic traveling waves which
are not ground states.
\end{remark}

\subsection{The Scaled Profile Equations and Fenichel Theory}
Traveling wave solutions
\begin{align}
u(x,t) = u(z) \hspace{1.0cm} z := x-ct
\end{align} to \eqref{g-RSPE}
must satisfy the singularly perturbed fourth order equation
\begin{equation}\label{profile_equations_almost}
\frac{d^2}{dz^2}\left(\b \frac{d^2u}{dz^2} + u^k -c u \right) + \a u
= 0
\end{equation} For any $\a, \b, c \neq 0$ consider the rescaling $u(z) = \tilde u (\tilde
z)$ where
\begin{equation}
\tilde u = |c|^{\frac{1}{1-k}}u \hspace{1.0cm} \tilde z =
\sqrt{\frac{|c|}{|\a|}} z
\end{equation} Then upon dropping tildes the profile equations \eqref{profile_equations_almost} become
\begin{equation}
\label{profile_equations_1} \frac{d^2}{dz^2}\left(\e^2 \frac{d^2
u}{dz^2} + {\rm sgn}(\b)u^k - {\rm sgn}(c \b) u \right) + {\rm
sgn}(\a \b) u = 0
\end{equation} where $\e$ is defined by \eqref{epsilon}. In order to use the GSPT framework we must rewrite the fourth order singulary perturbed
equation \eqref{profile_equations_1} in standard fast-slow form. In
doing this, a subtle point is the correct choice of slow and fast
variables, and a correct identification and placement of the small
parameter in the system of equations. The rewriting of the equation
as a system is not unique, and neither is the placement of the small
parameter in the equations, and hence it follows that the
identification of fast and slow variables is not unique either. With
this in mind, let $\e$ be as in \eqref{epsilon} and set
\begin{equation}\label{II:existence:fast_slow_variables}
u = u, \hspace{0.5cm} v = \e \dot u, \hspace{0.5cm} w = \e v + {\rm
sgn \b}u^k - {\rm sgn}(\b c)u, \hspace{0.5cm} y = \dot w
\end{equation} and consider $u , v$ the fast variables
and $w , y$ the slow variables. The motivation for this choice of
fast and slow variables \eqref{II:existence:fast_slow_variables}
will become clear in the subsequent analysis.  We may then write the
profile equations for traveling waves of \eqref{g-RSPE} as the
equivalent problem
\begin{equation}
\label{slow-system}
\begin{array}{rcl}
\e \dot{u} &=&  v \\
\e \dot{v} &=& w + {\rm sgn}(c \b) u - {\rm sgn}(\b) u^k \\
\dot{w} &=& -y  \\
\dot{y} &=& \rm{sgn}(\a \b) u
\end{array}
\end{equation}  where $\cdot$ denotes differentiation with respect to the slow variable $z$.  Define a fast variable $\z = z/\e$, then
\eqref{slow-system} becomes

\begin{equation}
\label{fast-system}
\begin{array}{rcl}
u^\prime &=& v \\
v^\prime &=& w + {\rm sgn}(c\b) u - {\rm sgn}(\b)u^k \\
w^\prime &=& -\e y  \\
y^\prime &=& \e {\rm sgn}(\a\b) u
\end{array}
\end{equation} where $\prime$ denotes differentiation with respect
to the fast variable $\z = z/\e$. We call \eqref{slow-system} the
{\em slow scaling} and \eqref{fast-system} the {\em fast scaling}.\\

We prove the existence of homoclinic orbits to \eqref{slow-system}
(or equivalently \eqref{fast-system}) via the Fenichel theory for
singularly perturbed systems of ordinary differential equations,
which forms the basis of geometric singular perturbation analysis.
We now briefly review the Fenichel theory.  For a very readable
account of this theory and its applications we refer to the C.I.M.E.
lectures by Jones
\cite{J} or Szmolyan \cite{S}.\\

Consider the system of autonomous ordinary differential equations
written in standard fast-slow form
\begin{equation}
\label{fast/slow}
\begin{array}{rcl}
\e \dot x &=& f(x,y;\e) \\
\dot y &=& g(x,y;\e)
\end{array}
\end{equation} where $x \in \R^n, y \in \R^m, f, \in C^\i$ and $0 < \e \leq \e_0 \ll 1$.  Let $z$ denote the independent variable in
(\ref{fast/slow}), which is referred to as the {\em slow scale}.  By introducing a {\em fast scale} $\z \equiv z/\e$ in \eqref{fast/slow},  one
obtains the equivalent system
\begin{equation}
\label{rescaled-fast/slow}
\begin{array}{rcr}
x' &=& f(x,y; \e) \\
y' &=& \e g(x,y;\e)
\end{array}
\end{equation} where $' \equiv \frac{d}{d\z}$.  The basic idea of GSPT is to analyze \eqref{fast/slow} by combining information derived from the {\em
reduced problem}

\begin{equation}
\label{reduced-fast/slow}
\begin{array}{rcl}
0 &=& f(x,y;0) \\
\dot y &=& g(x,y;0)
\end{array}
\end{equation} and the {\em layer problem}
\begin{equation}
\label{layer-fast/slow}
\begin{array}{rcl}
x' &=& f(x,y; 0) \\
y' &=& 0
\end{array}
\end{equation} which are the formal  $\e = 0$ limiting problems of (\ref{fast/slow}) and (\ref{rescaled-fast/slow}).  The fundamental connections
between the formal limiting problems and \eqref{fast/slow} were laid down by Fenichel in \cite{F} and consist of three main theorems referred to as
Fenichel's First, Second, and Third theorems.  The basic idea of the First Theorem is that \eqref{reduced-fast/slow} may be seen as a dynamical system
on the set ${\mathcal M} = \{ (x,y) \in \R^n \times \R^m : f(x,y;0) = 0 \}$ which is an invariant manifold of fixed points for
\eqref{layer-fast/slow}. Because of the trivial dynamics of $y' = 0$, ${\mathcal M}$ cannot be a hyperbolic set with respect to the full flow of
\eqref{layer-fast/slow}.  It can however be hyperbolic with respect to just the flow of $x' =f(x,y;0)$, which leads to the concept of {\em normal
hyperbolicity}.\\

\begin{definition}[Normal Hyperbolicity]
A manifold ${\mathcal M}$ is said to normally hyperbolic with
respect to a flow of the fast-slow system \eqref{fast/slow} if the
linearization of \eqref{fast/slow} at any point $\rho \in {\mathcal
M}$ when evaluated at $\e = 0$ has exactly $m$ zero eigenvalues and
$n$ eigenvalues  $\mu$ with $\Re \mu \neq 0$.
\end{definition} It turns out that this assumption is enough to prove the following persistence theorem of N. Fenichel, which guarantees that
${\mathcal M}$ persists as a manifold ${\mathcal M}_\e$ for $\e > 0$ small
enough.\\

\begin{theorem}[Fenichel's First Theorem \cite{F}, \cite{J}]
\label{Fenichel1} Suppose ${\mathcal M}$ is normally hyperbolic.
Then for $\e > 0$ but sufficiently small, there exists a manifold
${\mathcal M}_\e$ that lies within ${\mathcal O}(\e)$ of ${\mathcal
M}$ and is diffeomorphic to ${\mathcal M}$.  Moreover, it is locally
invariant under the flow of \eqref{fast/slow},
\eqref{rescaled-fast/slow}, and $C^k$, including in $\e$ for any $k<
+\i$.
\end{theorem}

Attached to each point $\rho \in {\mathcal M}$ there is a one
dimensional stable manifold ${\mathcal W}^s(\rho)$ and a one
dimensional unstable manifold ${\mathcal W}^s(\rho)$.  We collect
these manifolds together to form stable and unstable manifolds for
the full limiting slow manifold ${\mathcal M}$, given by

\begin{equation}
\begin{array}{rcl}
{\mathcal W}^s({\mathcal M}) &=& \bigcup_{\rho \in {\mathcal M}} {\mathcal W}^s(\rho) \\
{\mathcal W}^u({\mathcal M}) &=& \bigcup_{\rho \in {\mathcal M}}
{\mathcal W}^u(\rho)
\end{array}
\end{equation}   To address the question as to whether these
structures persist when $\e > 0$ we turn to Fenichel's Second
Theorem.\\

\begin{theorem}[Fenichel's Second Theorem \cite{F,J}]
Suppose ${\mathcal M}$ is normally hyperbolic.  Then for $\e>0$
sufficiently small, there exist a manifold ${\mathcal W}^s({\mathcal
M}_\e)$ (resp. ${\mathcal W}^u({\mathcal M}_\e)$) that lie within
${\mathcal O}(\e)$ of, and is diffeomorphic to,  ${\mathcal
W}^s({\mathcal M})$ (resp. ${\mathcal W}^u({\mathcal M})$).
Moreover, it is locally invariant under the flow of of
\eqref{fast/slow}, \eqref{rescaled-fast/slow} and $C^k$, including
in $\e$ for any $k< +\i$.
\end{theorem}

Moreover, there exist stable and unstable invariant foliations with
base ${\mathcal M}_\e$ with the dynamics along each foliation being
a small perturbation of a suitable restriction of the dynamics of
\eqref{layer-fast/slow}.\\

\begin{theorem}[Fenichel's Third Theorem \cite{F,J}]
Suppose ${\mathcal M}$ is normally hyperbolic.  Then for $\e > 0$
sufficiently small there exist for each point $\rho_\e \in {\mathcal
M}_\e$, manifolds that form an invariant family relative to
\eqref{fast/slow}, \eqref{rescaled-fast/slow}, which we name
${\mathcal W}^s(\rho)$ and ${\mathcal W}^u(\rho)$, which are within
${\mathcal O}(\e)$ of, and diffeomorphic to the corresponding
manifolds when $\e = 0$.
\end{theorem}

\subsection{Construction and Analysis of the Solitary Waves}

\subsubsection{The Slow Manifold and Reduced Problem} When $\e = 0$ in the
slow scaling \eqref{slow-system} we obtain
\begin{equation}
\label{reduced_problem}
\begin{array}{rcl}
0 &=& v \\
0 &=& w + {\rm sgn}(c\b) u - {\rm sgn}(\b)u^k \\
\dot{w} &=& -y  \\
\dot{y} &=& {\rm sgn}(\a\b ) u.
\end{array}
\end{equation}  Thus there is a manifold of fixed points for \eqref{reduced_problem} which we call the critical
manifold ${\mathcal M}_0$ given by
\begin{equation}
\label{M_0} {\mathcal M}_0 = \{ (u, v, w, y) \in \R^2\times I \times
\R : u = \psi(w), v=0 \}
\end{equation} where $\psi(w)$ is implicitly defined by the algebraic equation
\begin{equation}
\label{Psi} w  + {\rm sgn}(c\b) \psi - {\rm sgn}(\b)\psi^k   = 0
\end{equation} and
\begin{equation}\label{I} I := \{u : {\rm sgn}(\b)k u^{k-1}) < {\rm sgn}(c\b) \}
\end{equation}

This means that the flow on the limiting slow manifold $\cM_0$ is
given by
\begin{equation}
\label{reduced_system_e=0}
\begin{array}{rcl}
\dot{w} &=& -y  \\
\dot{y} &=& {\rm sgn}(\a\b) \psi(w).
\end{array}
\end{equation}

\begin{proposition}[Regime of Normal Hyperbolicity of ${\mathcal
M}_0$]\label{regime_of_normal_hyperbolicity} The critical manifold
${\mathcal M}_0$ defined through \eqref{M_0} is normally hyperbolic
with respect to the flow of \eqref{fast-system}.\\
\end{proposition}

\noindent {\bf Proof.} Setting $\e = 0 $ in the linearization of (\ref{fast-system})
at any point $\rho_0 \in \cM_0$ yields the matrix
\begin{equation}
\left( \begin{array}{ccccc}
0 & 1 & 0 & 0 \\
{\rm sgn}(c\b)-{\rm sgn}(\b)ku^{k-1} & 0 & 1 & 0  \\
0 & 0 & 0 & 0 \\
0 & 0 & 0 & 0  \\
\end{array} \right)
\end{equation}  which has characteristic polynomial
\[
\mu^2 \left(\mu^2 - {\rm sgn}(c\b) + {\rm sgn}(\b)ku^{k-1} \right) =
0.
\] Hence zero is an eigenvalue of multiplicity two, and there are two distinct eigenvalues
\[
\mu_\pm = \sqrt{{\rm sgn}(\b)({\rm sgn}(c)-ku^{k-1})}
\]  which have nonzero real part provided that
\begin{equation}
\label{normal_hyperbolicity_condition} {\rm sgn}(\b) ku^{k-1} < {\rm
sgn}(c\b).
\end{equation} However, by \eqref{I} every point in $\cM_0$
satisfies \eqref{normal_hyperbolicity_condition}.\\
\qed

\begin{remark}  Since $u$ is multivalued as a graph over $w$, the interval $I$ \eqref{I} is introduced to choose
the branch which contains the origin.\\
\end{remark}

\begin{proposition}[Existence and Characterization of ${\mathcal
M}_\e$] Under the hypothesis of Proposition
\eqref{regime_of_normal_hyperbolicity} there exists a manifold
${\mathcal M}_\e$ given by
\begin{equation}
\label{slow-manifold} {\mathcal M}_\e = {\mathcal M}_0 + {\mathcal
O}(\e^2)
\end{equation} and the flow on ${\mathcal M}_\e$ is given by

\begin{equation}
\label{reduced-problem} \begin{array}{rcl}
\dot w &=& -y  \\
\dot y &=& {\rm sgn}(\a \b)\psi(w) + {\mathcal O}(\e^2) \\
\end{array}
\end{equation}
\end{proposition}

\noindent {\bf Proof.} For the regime of normal hyperbolicity of $\cM_0$, Theorem
\ref{Fenichel1} guarantees the existence of a manifold ${\mathcal
M}_\e$ which is an ${\mathcal O}(\e)$ perturbation of $\cM_0$. A
straightforward calculation shows that $\cM_\e$ is in fact an
$\cO(\e^2)$ perturbation of $\cM_0$.
\qed

\begin{proposition}[Analysis of the Flow on $\cM_\e$]
The origin of the system \eqref{reduced_problem} is hyperbolic when
viewed as a dynamical system on ${\mathcal M}_\e$ if and only
if
\begin{equation}\label{a/c}
{\rm sgn}(\a) = {\rm sgn}(c) .
\end{equation}  Furthermore, when \eqref{a/c} holds the origin has a one-dimensional stable and
one-dimensional unstable manifold with eigenvectors of the
linearization of \eqref{reduced-problem} at the origin begin
$\cO(\e^2)$ perturbations of
\begin{equation}
 \mu_0 = 0  \;\; (\text{double eigenvalue}) \hspace{1.0cm} \mu_\pm =
\pm 1
\end{equation}
\begin{equation}\label{reduced_problem_eigenvalues}
\eta_0 = \left( \begin{array}{c} 0 \\ 0 \\ \ast \\ \ast \end{array}
\right) \hspace{1.0cm} \eta_- = \left( \begin{array}{c} \ast \\ \ast
\\ -1 \\ {\rm sgn}(\a\b)
\end{array} \right) \hspace{1.0cm} \eta_+ = \left( \begin{array}{c}
\ast \\ \ast \\ 1 \\ {\rm sgn}(\a\b) \\
\end{array} \right)
\end{equation}
\end{proposition}

\noindent {\bf Proof.} First note that the origin is an element of $\cM_0$. Since the origin
is a fixed point for the full problem \eqref{slow-system} it remains
a fixed point for the flow on $\cM_\e$ also.  To establish
hyperbolicity of the origin as well as the eigenvectors of the
linearized flow, we need only establish them for the limiting
problem \eqref{reduced_problem} or \eqref{reduced_system_e=0} which
are $\mu = \pm \sqrt{-\psi'(0)}$. By differentiating \eqref{Psi} we
get $\psi'(0) = -{\rm sgn}(\a \b) / {\rm sgn}(c \b)$, so that $\mu =
\pm \sqrt{{\rm sgn}(\a){\rm sgn}(c)}$ and the result follows.\\
\qed

\subsubsection{Analysis of the Layer Problem}
We now consider the fast scaling \eqref{fast-system}. Setting $\e =
0$ we have the layer problem
\begin{equation}
\label{layer-problem} \left.
\begin{array}{rcl}
u^\prime &=& v \\
v^\prime &=& w + {\rm sgn}(c\b) u - {\rm sgn}(\b)u^k \\
w^\prime &=& 0 \\
y^\prime &=& 0 \\
\end{array} \right\}
\end{equation}
Essential in the construction of the traveling wave is that the
layer equation has an orbit that connects points on the critical
manifold ${\mathcal M}_0$. By the definition of $\cM_0$ \eqref{M_0},
this means the fixed points satisfy
\begin{equation}
w =  {\rm sgn}(\b)u_\ast^k - {\rm sgn}(c\b)u_\ast.
\end{equation}  However, the fixed points of the layer problem must
remain fixed points of the full system, and since we are looking for
solutions homoclinic to the origin, we require $w=0$ which implies
\begin{equation}\label{speed}
u_\ast^{k-1} = {\rm sgn}(c).
\end{equation}  Note that if $k$ is odd the speed $c$ must be
positive. Thus \eqref{layer-problem} becomes
\begin{equation}
\label{layer} \left. \begin{array}{rcl}
u^\prime &=&   v\\
v^\prime &=& {\rm sgn}(c\b) u - {\rm sgn}(\b)u^k  \\
w^\prime &=& 0 \\
y^\prime &=& 0
\end{array} \right\}
\end{equation}  If $k$ is odd the fixed points of
\eqref{layer} are $(-1, 0,0,0)$, $(0,0,0,0)$, $(1,0,0,0)$ while if
$k$ is even the fixed points of \eqref{layer} are $(0,0,0,0)$ and
$({\rm sgn}(c), 0,0,0)$.

\begin{proposition}[Analysis of the Layer Problem]
\label{II:prop:layer} The layer problem \eqref{layer} has an orbit
homoclinic to the origin if and only if
\begin{align}\label{c/B}
{\rm sgn}(\b) = {\rm sgn}(c) \, .
\end{align}  The orbit is given by
\begin{align}\label{homoclinic}
h(\z) = (q(\z), - q(\z) {\rm tanh}\left(\frac{k-1}{2}\z \right),0,0)
\end{align} where
\begin{align}\label{gKdV_soliton}
q(\z) = {\rm sgn}(c)\left(\frac{(k+1)}{2} \right)^{\frac{1}{k-1}}\,
{\rm sech}^{\frac{2}{k-1}}\left(\frac{k-1}{2}\z \right)
\end{align} and ${\rm sgn}(c) = \pm 1$ when $k$ is even while ${\rm sgn}(c) =
1$ when $k$ is odd.
\end{proposition}

\noindent {\bf Proof.} Since the center directions play no role in the construction
of the homoclinic orbit we can restrict the flow of \eqref{layer} to
the two-dimensional $(u,v)$ phase space
\begin{equation}\label{profile-gKdV}
\left. \begin{array}{rcl}
u^\prime &=& v \\
v^\prime &=& {\rm sgn}(c\b) u - {\rm sgn}(\b)u^k \\
\end{array} \right\}
\end{equation}  This equation is the profile of the generalized KdV (gKdV)
solitary wave, which has the well known homoclinic solution
\eqref{gKdV_soliton} provided the sign of $c$ and $\b$ are the same.
\qed

\begin{remark}\label{rem:parameter-regime}
Split the parameter space into $\bP=\bP_-\cup \bP_+$ where $\bP_\pm
:= \{(\a,\b,c) \in \R^3 : {\rm sgn}(\a) = {\rm sgn}(\b) = {\rm
sgn}(c) = \pm 1, \e \ll 1\}$. From \eqref{speed} we see that for $k$
odd, ${\rm sgn}(c)=1$ so that \eqref{c/B} implies ${\rm sgn}(\b)=1$
and \eqref{a/c} implies ${\rm sgn}(\a)=1$. Thus for $k$ odd we can
construct solitary waves in the parameter space $\bP_+$ while for
$k$ even we can construct solitary waves in $\bP=\bP_-\cup \bP_+$.
Without loss of generality, in what follows we assume $(\a,\b,c) \in
\bP$
\end{remark}

\subsection{Tangent Spaces and the Transversality Calculation} Now that
we have analyzed the $\e = 0$ limiting systems in both the slow and
fast scaling, we prove the existence of solitary waves to the full
$\e>0$ problem. We prove this via a reversibility argument  together
with a transversality calculation, which essentially entails showing
that certain manifolds associated with the slow and fast orbits are
transverse at $\e=0$ and thus transverse for $\e>0$ small
enough.\\

\subsubsection{The Reversibility Argument} Here we formulate
and verify a condition that, together with the transversality
calculation in Section \ref{sect:transversality} below, proves the
existence of a homoclinic orbit of the profile equations for the
RSPE equations. This condition
essentially follows from reversability of the dynamical system.\\

\begin{proposition}[Condition for the Existence of a Homoclinic
Orbit] \label{transversality_condition} Suppose $\G(z) : \R
\rightarrow \R^4$ is an orbit of \eqref{slow-system} (or
equivalently \eqref{fast-system}) which satisfies
\begin{align}\label{trans_1}
\lim_{z\to -\i} \G(z) = 0
\end{align}  and for some $z=z^\ast$,
\begin{align} \label{trans_2}
\Pi \cap \G(z^\ast) \; \mbox{is nonempty}.
\end{align}  Then $\G(z)$ is a homoclinic orbit.
\end{proposition}

\noindent {\bf Proof.} First notice without loss of generality the translational
invariance allows us to set $z^\ast=0$. Next, since the profile
equations in both the slow scaling \eqref{slow-system} and the fast
scaling \eqref{fast-system} are invariant under the transformation
\begin{equation}
\label{symmetry_mapping} z \mapsto -z \hspace{1.0cm} (u,v,w,y)
\mapsto (u,-v,w,-y)
\end{equation} there exists a reversibility operator $\cR(z)$
such that
\begin{equation}
\cR(u,v,w,y)(z) = \cR(u,-v,w,-y)(-z)
\end{equation} with $\mbox{Fix} \cR$ the two dimensional plane
\begin{align}
\Pi :=\{(u,v,w,y)\in\R^4 : v=0, \; y=0 \} .
\end{align}  One may take $\cR$ as
\begin{equation}
\cR = \left( \begin{array}{cccc} 1 & 0 & 0 & 0 \\
0 & -1 & 0 & 0 \\
0 & 0 & 1 & 0 \\
0 & 0 & 0 & -1
\end{array} \right).
\end{equation} Recall that the origin of \eqref{slow-system},
\eqref{fast-system} is a hyperbolic fixed point with two stable and
two unstable eigenvalues. Clearly then there is a $\G(z)$ such that
$\G$ satisfies \eqref{trans_1}.  Suppose that in addition it
satisfies \eqref{trans_2} at the point $z = 0$. This means we have
constructed $\{ \G (z) : -\i \leq z \leq 0 \}$. By applying $\cR$ to
this portion of $\G$ we can construct $\{ \G(z) : 0 \leq z \leq
\i\}$. By reversibility we have that $\G(z) = \cR \G(-z)$ so that
\begin{align}
\begin{split}
\lim_{z \to \i} \G(z) & = \lim_{z \to \i} \cR \G(-z) \\
& = \cR \lim_{z \to \i} \G(-z) \\
& = 0
\end{split}
\end{align} which by definition means $\G(z)$ is a homoclinic
orbit. \qed 

By Proposition \ref{transversality_condition}, all that is left to
prove in order to establish the existence of a homoclinic orbit to
\eqref{slow-system}, \eqref{fast-system} is that for $\e>0$ small
enough there is an orbit that satisfies \eqref{trans_1} and
\eqref{trans_2}.

\subsubsection{The Transversality Calculation}\label{sect:transversality} Consider the profile
equations in the fast scaling \eqref{fast-system}.  When $\e=0$,
\eqref{fast-system} has a homoclinic orbit $Q= (q,q',0,0)$ which
satisfies \eqref{trans_1} and \eqref{trans_2}. Showing that this
holds also for $\e >0$ small enough amounts to showing that the
conditions for the implicit function theorem hold. This in turn
amounts to showing that the evolution of the two dimensional
unstable manifold under the flow of
\eqref{fast-system} when $\e = 0$ projected onto the orthogonal complement of $\Pi$, $\Pi^\perp$, is nonzero.\\

Consider the equations in the fast scaling \eqref{fast-system} in
differential form notation,
\begin{equation}\label{diff_form_fast}
\left.
\begin{array}{rcl}
du^\prime &=& dv \\
dv^\prime &=& dw + (1-{\rm sgn}(c)ku^{k-1})du  \\
dw^\prime &=& -\e dy \\
dy^\prime &=& \e du .  \\
\end{array} \right\}
\end{equation}  where we have used Remark \ref{rem:parameter-regime} to simplify the equations. Setting $\e = 0$ in \eqref{diff_form_fast} yields
\begin{equation}\label{diff_form_fast_0}
\left.
\begin{array}{rcl}
du^\prime &=& dv \\
dv^\prime &=& dw + (1-{\rm sgn}(c)ku^{k-1})du  \\
dw^\prime &=& 0 \\
dy^\prime &=& 0   \\
\end{array} \right\}\, .
\end{equation}

Consider the two form
\begin{equation}
dv(\z) \wedge dy(\z) \in \Lambda^2 \R^4 \, .
\end{equation}  Recall that the two dimensional critical manifold
$\cM_0$ is given by
\begin{equation}
\cM_0 := \{(u,v,w,y) \in \R^2\times I \times \R : u = \psi(w), \; v
= 0 \} .
\end{equation}  By the Fenichel theory, the unstable (resp. stable) manifold of
$\cM_\e$ is completely foliated by smooth curves referred to as
Fenichel fibers. Each Fenichel fiber intersects $\cM_0$ at a unique
point called the basepoint of the fiber. Thus, the foliation is a
2-parameter family of one-dimensional curves. The important feature
of these fibers is that points on a fiber correspond to initial
conditions that asymptotically approach the orbit on $\cM_0$ as $z
\rightarrow \i$ (resp. $z \rightarrow \i$) that passes through the
basepoint of that particular fiber. Let $f^u(\rho_0)$ denote an
unstable fiber contained in $\cW^u(\cM_0)$ which has basepoint
$\rho_0 \in \cM_0$ and $f^s(\rho_0)$ denote a stable fiber contained
in $\cW^s(\cM_0)$ which has basepoint $\rho_0 \in \cM_0$. The
Fenichel theory enables us to identify lower-dimensional invariant
manifolds within these stable and unstable manifolds.  Let $\g_0
\subset \cM_0$ be an orbit on the slow manifold $\cM_0$ satisfying
$\lim_{z\rightarrow -\i}\g_0 = 0$, then it has its own unstable
manifold, denoted by $\cW^u(\g_0)$, which is simply the union of all
unstable fibers which have their basepoints lying on $\g_0$.  We now proceeds with the calculation.\\

Let $\eta_1$ be a vector tangent to the reduced problem which we
take to be
\begin{align}
\eta_1 = (\psi'(w),0,-1,{\rm sgn}(\a)) \in T \cW^u \, .
\end{align} Since $\lim_{\z \rightarrow \pm \i} w(\z) = 0$ then by
continuity
\begin{align}
\lim_{\z \rightarrow \pm \i} \psi'(w) = \psi'(0) = -1
\end{align} and we can take
\begin{equation}
\label{II:eta_1} \eta_1 = (-1,0,1,{\rm sgn}(c)).
\end{equation} By definition, $\eta_1$ is tangent to the reduced
flow at $z = -\i$.  Let $\Phi^z  = (\vp_1^z, \vp^z_2, \vp^z_3,
\vp^z_4)$ denote the flow of \eqref{diff_form_fast_0}.  Then $\g_0
:= \Phi^z \cdot \eta_1$ defines an orbit in $\cM_0$.  Notice that
for the layer problem \eqref{diff_form_fast_0}, for every $\z\in \R$
there is no flow for both $w$ and $y$ so that the third and fourth
component $\eta_1$ is invariant under $\Phi^z$, that is $\Phi^z
\cdot \eta_1 = -\vp^z_1
\cdot (1 , 0 ,1,{\rm sgn}(\a))$.\\

Next take a vector tangent to the $\e=0$ homoclinic orbit $h$
\eqref{homoclinic}, which we take as the vector field of the flow in
the fast scaling at $\e=0$, \eqref{layer-problem}
\begin{equation}
\label{II:eta2} \eta_2 = (v, w + u - {\rm sgn}(c)u^k, 0, 0).
\end{equation} Clearly, \eqref{II:eta2} gives a vector tangent to the homoclinic orbit $h$ for every $z \in
\R$ and furthermore, $\eta_2 \in T \cW^u$.  We now compute the
projection of the limiting $\e=0$ fast flow onto $\Pi$ at $z=0$ when
applied to the the tangent space of $\cW^u$. That is, we wish to
compute
\begin{equation}
(dv \wedge dy)(\eta_1, \eta_2)(0) \, .
\end{equation}

To do this notice that
\begin{align}
\begin{split}
\frac{d}{d\z} \left(dv \wedge dy \right)(\eta_1,\eta_2)(\z) &= (dv'
\wedge dy)(\eta_1,\eta_2)(\z) + (dv \wedge dy')(\eta_1,\eta_2)(\z) \\
& = (1-{\rm sgn}(c)ku^{k-1})(du \wedge
dy)(\eta_1,\eta_2)(\z) + (dv \wedge 0)(\eta_1,\eta_2)(\z)\\
&= (1-{\rm sgn}(c)ku^{k-1})(du \wedge
dy)(\eta_1,\eta_2)(\z) \\
&= -(1-{\rm sgn}(c)ku^{k-1})v \\
&= -(1-{\rm sgn}(c)ku^{k-1})u' \, .
\end{split}
\end{align}  Thus
\begin{align}\label{trans_calc}
\begin{split}
(dv \wedge dy)(\eta_1, \eta_2)(0) &= \int_{-\i}^0 \frac{d}{d\z}
\left(dv \wedge dy \right)(\eta_1,\eta_2)(\z) d\z \\
&= - \int_{-\i}^0 (1-{\rm sgn}(c)kq^{k-1})q' \, d\z \\
&= - \left(q(0) - {\rm sgn}(c) q(0)^{k}  \right) \\
&= {\rm sgn}(c)\left(\frac{k-1}{k+1}  \right) \left(\frac{k+1}{2}
\right)^{\frac{k}{k-1}}
\end{split}
\end{align} which is nonzero since by assumption $\a , c$ nonzero.  Since $\cW^u_\e$ and $\cW^s_\e$ intersect when $\e=0$ \eqref{trans_calc} shows
that this intersection is transverse.
Therefore, by the implicit function theorem, the manifolds still
intersect for $\e >0$ small enough. The intersection of $\cW^u_\e$
and $\cW^s_\e$ for nonzero $\e$ finishes the
construction of the pulse. \\
\qed

\subsection{The Melnikov Calculation, Homoclinic Breaking, and
Asymptotic Decay of the Wave} Here we want to prove some analytic
and geometric properties of the wave for $|z|$ large.  In particular
we want to show that the homoclinic orbit that exists when $\e>0$
enters (resp. exits) tangent to the weakly stable (resp. unstable)
eigenvectors. The idea is to show that when $\e>0$ is small but
nonzero, the homoclinic orbit \eqref{homoclinic} of the $\e=0 $
layer problem breaks. That is, the one dimensional stable and
unstable manifolds of the origin which intersect for the $\e=0$
layer problem, fail to intersect when $\e
> 0$. Thus the homoclinic orbit of the RSPE equations cannot enter (resp. exit) tangent to the eigenvectors associated with the
linearization of the layer problem at the origin.  This will be
proved via a Melnikov calculation. Using the fact that the profile
equations for RSPE are reversible, this means that the solution must
enter (resp. exit) the origin tangent to the eigenvectors associated
with the linearization of the reduced problem at the origin.

\subsubsection{The Melnikov Integral Calculation}  Here we show that
the homoclinic orbit that exists in the fast scaling when $\e = 0$
breaks when $\e > 0$.  To set up for the calculation, write the
traveling wave equations in the fast scaling
\eqref{profile_equations_1} as
\begin{equation}
\label{fast-system-2}
\begin{array}{rcl}
u^\prime &=& v\\
v^\prime &=& w + u - {\rm sgn}(c)u^k \\
w^\prime &=& y  \\
y^\prime &=& - \e^2 u \,.
\end{array}
\end{equation} Notice here that once again the issue of the correct
placement of the parameter $\e$ is important.  Set $U=(u,v,w,y)$ and
write \eqref{fast-system-2} as
\begin{align}
U' = F(U;\e^2)
\end{align} with $F$ defined by the right hand side of
\eqref{fast-system-2}.  When $\e = 0$ we have shown in Proposition
\ref{II:prop:layer} that the equations posses a homoclinic orbit
$h=(q,q',0,0)$ \eqref{homoclinic}. By reversibility of the profile
equations for the layer problem (also setting $\e = 0$ in
\eqref{fast-system-2}) we see that the one dimensional stable and
unstable manifolds for the layer problem intersect in the $(u,v)$
plane along $v=0$.  One way to measure how much the stable and
unstable manifolds of the the layer problem miss each other is to
define the distance between these curves evaluated along $v=0$ which
to first order is given by the Melnikov integral
\eqref{Melnikov_integral} which we now describe.\\

Consider the variational equations obtained by linearizing about the
$\e=0$ homoclinic orbit $h=(q,q',0,0)$ \eqref{homoclinic}, given by
\begin{align}
V' = D_U F(h;0) V
\end{align} which we write explicitly as
\begin{equation}
\label{variational-q}
V' = \left( \begin{array}{cccc} 0 & 1 & 0 & 0 \\
1-{\rm sgn}(c) kq^{k-1} & 0 & 1 & 0 \\
0 & 0 & 0 & 1\\
0 & 0 & 0 & 0
\end{array} \right) V \, .
\end{equation}  The adjoint variational equations
\begin{equation}
\Psi ' = -D_U^\dagger F(h;0) \Psi
\end{equation} are given explicitly by the system
\begin{equation}
\label{adjoint-variational-q}
\Psi' = \left( \begin{array}{cccc} 0 & -(1-{\rm sgn}(c)kq^{k-1}) & 0 & 0 \\
-1 & 0 & 1 & 0 \\
0 & -1 & 0 & 0\\
0 & 0 & -1 & 0
\end{array} \right) \Psi \, .
\end{equation}  where $\Psi = (\psi_1,\psi_2,\psi_3,\psi_4)$. From
\eqref{adjoint-variational-q} we see that $\psi_2$ satisfies the
equation
\begin{align}\label{psi_2}
\psi_2'' = 1-{\rm sgn}(c)kq^{k-1}\psi_2
\end{align} so that $\psi_2 = q'$ solves \eqref{psi_2} since it yields the profile equations for the gKdV equation \eqref{profile-gKdV}.
Since $\psi_1 = -\psi_2 '$, $\psi_3' = -\psi_2$ and $\psi_4' =
-\psi_3$,
\begin{equation}
\Psi = \left(
\begin{array}{c}
\psi_1 \\
\psi_2 \\
\psi_3 \\
\psi_4
\end{array} \right) = \left(
\begin{array}{c}
-q'' \\
q' \\
-q \\
\int_{-\i}^\z q \, d\z
\end{array} \right) \, .
\end{equation}

Let
\begin{equation}\label{Melnikov_integral}
M = \int_{-\i}^\i \<D_{\e^2} \left. F(h(\z);\e^2)\right|_{\e=0} ,
\Psi(\z) \>
\end{equation} where here $\< \; \cdot \; , \; \cdot \; \> $ denotes
the vector inner product, and $F(h;0)$ is the vector field defined
by setting $\e=0$ in the right hand side of \eqref{fast-system-2}.
Note that while $D_U F(U;0)$ is not of full rank and thus the origin
of \eqref{fast-system-2} is not hyperbolic, $M$ defined by
\eqref{Melnikov_integral} nonetheless defines a Melnikov integral in
the usual sense (c.f. \cite{CL}). Thus the homoclinic orbit breaks
for $\e>0$ if $M \neq 0$. We now evaluate the Melnikov integral \eqref{Melnikov_integral}.\\

Clearly
\begin{equation}
\left. D_{\e^2} F(U;\e^2) \right|_{\e = 0} = \left(\begin{array}{c}
0 \\
0 \\
0 \\
q
\end{array} \right)
\end{equation} so we have
\begin{align}
\begin{split}
M & = \int_{-\i}^\i \<D_{\e^2} \left. F(h(\z);\e^2)\right|_{\e=0} ,
\Psi(\z) \> \\
& = - \int_{-\i}^\i q \psi_4 \, d\z \\
&= -\int_{-\i}^\i q \left( \int_{-\i}^\z q \,dz \right) \, d\z \\
&= -2 \left(\int_0^\i q d\z  \right)^2
\end{split}
\end{align}  Since $q \geq 0$ for all $\z \in \R$ we have
\begin{align}
M \neq 0
\end{align} which proves the result.\\

\begin{remark}  The geometric significance of the Melnikov
calculation is that the homoclinic orbit of the full $\e > 0$
problem cannot enter (resp. exit) the origin tangent to the strongly
stable (resp. strongly unstable) eigenvectors.  This situation is
evidence of an {\em orbit-flip bifurcation} in $\e$, and allows us
to construct multi-bump traveling waves \cite{CJMS}.
\end{remark}

\begin{proposition}[Asymptotic Decay of the Wave.] Assume $\e>0$ fixed and $u(z)$ is a solitary wave of \eqref{g-RSPE}.  Let
\begin{equation}
\mu_\e = |\mu^{ws}_\e| = \mu^{wu}_\e = \frac{1}{2}\sqrt{2{\rm
sgn}(c) - 2\sqrt{1 - 4 \e^2}} \, .
\end{equation} Then for any $m \geq 0$ there exists an $R>0$ large enough so that
\begin{equation}
|\partial_z^m u(z)| \leq C_{\e}e^{-\mu_\e |z|}
\end{equation} for $|z|> R$.
\end{proposition}

 \noindent {\bf Proof.} The Melnikov calculation coupled with the reversibility argument
means that the homoclinic orbit of
\eqref{slow-system}\eqref{fast-system} which we constructed for
$\e>0$ small enough cannot enter (resp. exit) tangent to the fast
directions, and so must enter (resp. exit) tangent to the slow
directions. The eigenvectors associated to the slow directions have
magnitude $\mu_\e$. \qed


\begin{thebibliography}{CMJ1}
\bibitem{AC}{\sc M. Ablowitz \& P. Clarkson}, {\em Solitons, Nonlinear Evolution Equations and Inverse Scattering},
Cambridge University Press, Cambridge, 1991. \bibitem{BFRW} {\sc P.
W. Bates, P. C. Fife, X. Ren, \& X. Wang}, {\em Traveling Waves in a
Convolution Model for Phase Transitions}, Archive for Rational
Mechanics and Analysis, {\bf 138}, Number 2 (1997).
\bibitem{B}{\sc R.W. Boyd}, {\em Nonlinear Optics}, Academic Press, Boston, 1992.
\bibitem{CL}{\sc S.-N. Chow, \& X.-B. Lin}, {\em Bifurcation of a homoclinic orbit
with a saddle-node equilibrium}, Differential Integral Equations
{\bf 3}, (1990), no. 3, 435--466.
\bibitem{CIL}{\sc R. Choudhury, R.I. Ivanov, Y. Liu} {\em Hamiltonian formulation, nonintegrability, and local bifurcation for the Ostrovsky
equation}, Chaos, Solitons, and Fractals, {\bf 34}, No. 2, (2007),
544--550.
\bibitem{CJMS}{\sc N. Costanzino, C.K.R.T. Jones, V. Manukian, \& B.
Sandstede}, Existence of multiple-bump traveling waves of the regularized short pulse equation, Preprint.
\bibitem{CJSW} {\sc Y. Chung, C.K.R.T. Jones, T. Sch\"afer \& C.E. Wayne}, {\em Ultra-short pulses in linear and nonlinear media}, Nonlinearity, {\bf
    18}, (2005), 1351--1374.
\bibitem{CF}{\sc R. Courant \& K.O. Freidrichs}, {\em Supersonic flow and shock waves}, Springer-Verlag, New York, 1976.
\bibitem{Daf}{\sc C. Dafermos}, {\em Hyberbolic Conservation Laws in Continuum Physics}, Springer-Verlag, 2000.
\bibitem{GGS}{\sc O.A. Gilman, R. Grimshaw \& Yu. A. Stepanyants}, {\em Approximate and numerical solutions of the stationary Ostrovsky equation},
    Stud. Appl. Math., {\bf
95}, (1995), 115--126.
\bibitem{F} {\sc N. Fenichel}, {\em Geometric singular perturbation theory for ordinary differential equations}, J. Diff. Eq., {\bf 31}, (1979),
    53--98.
\bibitem{Fi}{\sc P.C. Fife}, {\em Travelling waves for a nonlocal double-obstacle
problem}, European Journal of Applied Mathematics, {\bf 8}, (1997),
581-594.
\bibitem{Gr}{\sc R. Grimshaw}, {\em Evolution equations for weakly
nonlinear long internal waves in a rotating fluid}, Stud. Appl.
Math., {\bf 73}, (1985), 1--33.
\bibitem{Hun}{\sc J.K. Hunter}, {\em Numerical solutions of some
nonlinear dispersive wave equations}, Lectures in Appl. Math., {\bf
26}, (1990), 301--316
\bibitem{J} {\sc C.K.R.T. Jones}, {\em Geometric singular perturbation theory}, in {\em Dynamical systems}: Lecture Notes in Math., {\bf 1609},
    Springer-Verlag, Berlin-New York, (1994), 44--118.
\bibitem{KNNSMSY}{\sc N. Karasawa, S. Nakamura, N. Nakagawa, M. Shibata, R.
Morita, H. Shigekawa, \& M. Yamashita}, {\em Comparision between
theory and experiment of nonlinear propagation for a-fewcycle and
ultrabroadband optical pulses in a fused-silica fiber}, IEEE J.
Quant. Elect., {\bf 37}, (2001), 398–-404.
\bibitem{LL}{\sc S. Levandovsky \& Y. Liu}, {\em Stability of solitary waves of a generalized Ostrovsky equation}, SIAM J. Math. Anal., {\bf 38},
    (2006), 985-1011.
\bibitem{LV}{\sc Y. Liu \& V. Varlamov}, {\em Stability of solitary waves and weak rotation limit
for the Ostrovsky equation}, J. Differential Equations, {\bf 203},
(2004), 159–-183.
\bibitem{M}{\sc I. H. Maliton}, {\em Interspecimen comparison of the refractive index of fused silica}, J. Opt. Soc. Amer., {\bf 55}, October, (1965),
    1205--1210.
\bibitem{MN}{\sc B. P. Marchant \& John Norbury}, {\em Discontinuous travelling wave solutions for certain hyperbolic
systems}, IMA Journal of Applied Mathematics, {\bf 67}, (2002),
201-224.
\bibitem{NM} {\sc A.C. Newell \& J.V. Moloney}, {\em Nonlinear Optics}. Addison-Wesley, Redwood City, CA, (1992).
\bibitem{NSC}{\sc S. P. Nikitenkova, Yu. A. Stepanyants \& L. M.
Chikhladze}, {\em Solutions of the modified Ostrovskii equation with
cubic non-linearity}, J. Appl. Maths Mechs, {\bf 64}, No. 2, (2000),
267--274.
\bibitem{O} {\sc L.A. Ostrovsky}, {\em Nonlinear internal waves in a rotating
ocean}, Okeanologia, {\bf 18}, 2 (1978), 181–-191.
\bibitem{Ro}{\sc J. E. Rothenberg}, {\em Space-time focusing:
breakdown of the slowly varying envelope approximation in the
self-focusing of femtosecond pulses}, Opt. Lett., {\bf 17}, (1992),
1340–-1342.
\bibitem{SS} {\sc A. Sakovich \& S. Sakovich}, {\em The short pulse
equation is integrable}, J. Phys. Soc. Jpn., {\bf 74}, (2005),
239--241.
\bibitem{S}{\sc P. Szmolyan}, {\em Transversal heteroclinic and homoclinic orbits in singular perturbation problems}, J. Differential Equations, {\bf
    92} (1991), no. 2, 252--281.
\bibitem{SW} {\sc T. Sch\"afer \& C.E. Wayne}, {\em Propagation of ultra-short optical pulses in cubic nonlinear media}, Phys. D, {\bf
196}, (2004), 90 -- 105.
\end{thebibliography}
\end{document}